\documentclass[journal]{IEEEtran}
\ifCLASSINFOpdf
\else
\fi
\usepackage[cmex10]{amsmath}
\newtheorem{theorem}{Theorem}[section]
\newcommand{\bew}{\proof}
\newtheorem{rem}[theorem]{Remark}
\newtheorem{example}[theorem]{Example}
\newtheorem{lemma}[theorem]{Lemma}
\newtheorem{definition}[theorem]{Definition}
\newtheorem{cor}[theorem]{Corollary}

\DeclareMathOperator{\wt}{wt}
\DeclareMathOperator{\wtfive}{wt_5}

\DeclareMathOperator{\diag}{diag}

\DeclareMathOperator{\GL}{GL}
\DeclareMathOperator{\Aut}{Aut}

\newcommand{\disj}{\stackrel{.}{\cup}}

\newcommand{\F}{{\mathbf F}}

\newcommand{\Z}{{\mathbf Z}}
\newcommand{\trace}{\mbox{trace}}

\newcommand{\eb}{\endproof}

\begin{document}
%
\title{
The automorphism group of an extremal $[72,36,16]$ code does not contain
$Z_7$, $Z_3\times Z_3$, or $D_{10}$.}
%
%
%

\author{Thomas Feulner and  Gabriele Nebe%
\thanks{T. Feulner is with the 
Mathematics Department, University of Bayreuth, 95440 Bayreuth, Germany, 
e-mail: thomas.feulner@uni-bayreuth.de --  supported by DFG priority program SPP 1489}%
\thanks{G. Nebe is with the Lehrstuhl D f\"ur Mathematik, RWTH Aachen University,
52056 Aachen, Germany, e-mail: 
 nebe@math.rwth-aachen.de.}
\thanks{Manuscript received October, 2011; revised .}}

%
%

\markboth{IEEE Transactions on Information Theory, Vol. IT-XX, No. X, Month 2011}%
{Feulner and Nebe: The automorphism group of an extremal $[72,36,16]$ code does not contain
$Z_7$, $Z_3\times Z_3$, or $D_{10}$.}
%



\maketitle

\begin{abstract}
A computer calculation with Magma shows that there is no
extremal self-dual binary code $C$ of length $72$ that has an
automorphism group containing either the dihedral group $D_{10}$ of order 10,
the elementary abelian group $Z_3\times Z_3$ of order 9, or the cyclic 
group  $Z_7$ of order 7.
Combining this with the known results in the literature one obtains that 
$\Aut(C)$ is either $Z_5$ or has order dividing $24$. 
\end{abstract}


%
\IEEEpeerreviewmaketitle

\section{Introduction}
%
%
%
%
\IEEEPARstart{L}{et} 
$C=C^{\perp }\leq \F_2^n$ be a binary self-dual code of length $n$.
Then all weights 
$\wt(c) := |\{ i \mid c_i = 1 \} | $
of codewords in $C$ are even and $C$ is called
{\em doubly-even}, if
$\wt(C) :=\{ \wt(c) \mid c\in C \} \subseteq 4\Z$. 
Doubly-even self-dual binary codes are also called {\em Type II} codes.
Using invariant theory, one may show \cite{MallowsSloane}
that the minimum weight
$d(C):= \min (\wt (C\setminus \{ 0 \} )) $ of a Type II code is bounded from above
by $$d(C) \leq 4 + 4 \lfloor \frac{n}{24} \rfloor .$$
Type II codes achieving this bound are called {\em extremal}.
Particularly interesting are the extremal codes of length a multiple of
24.
There are unique extremal codes of length 24 (the
extended binary Golay code ${\mathcal G}_{24}$) and 48
(the extended quadratic residue code QR$_{48}$), and both
have a fairly big automorphism group (namely $\Aut({\mathcal G}_{24}) 
\cong M_{24}$ and $\Aut($QR$_{48}) \cong $ PSL$_2(47)$) acting at least
2-transitively.
The existence of an extremal code of length 72  is a longstanding
open problem (see \cite{sloane}).
A series of papers investigates the automorphism group of a
putative extremal code of length 72
excluding most of the subgroups of $S_{72}$.
Continuing these investigations we show the following theorem, 
which is the main result of this paper:

\begin{theorem} \label{main}
The automorphism group of a binary self-dual doubly-even
$[72,36,16]$ code has order $5$ or $d$ where
$d$ divides $24$.
\end{theorem}

Throughout the paper the {\em cyclic group} of order $n$ is denoted by $Z_n$ to reserve the 
letter $C$ for codes. With $D_{2n}$ we denote the {\em dihedral group} of order $2n$, $S_n$ and $A_n$ 
are the {\em symmetric} and {\em alternating} groups of degree $n$.
$G\times H$  denotes the {\em direct product} of the two groups $G$ and $H$ and
 $G\wr S_{n} $ denotes the {\em wreath product} with normal subgroup 
$G\times G \ldots \times G$ and the symmetric group of degree $n$ 
permuting the $n$ components.

The following is known about the automorphism group $\Aut(C)$ of a 
binary self-dual doubly-even
$[72,36,16]$ code $C$:

By \cite[Theorem 1]{BW} 
the group $\Aut(C)$ has order $5$, $7$, $10$, $14$, or $d$ where 
$d$ divides $18$ or $24$ or $\Aut(C) \cong A_4\times Z_3$. 
The paper 
 \cite{Yankov} shows that $\Aut(C)$ contains no element of order $9$, 
 \cite[Corollary 3.6]{NebeF2}
excludes $Z_{10}$ as subgroup of $\Aut(C)$. 
So to prove Theorem \ref{main} 
it suffices to show that there are no such codes $C$ for which $\Aut(C)$ contains
 $D_{10}$ (Theorem \ref{main3}), $Z_7$ (Theorem \ref{main2}), or $Z_3\times Z_3$ (Theorem \ref{main1}).
The necessary computations, which have been performed in Magma \cite{MAGMA}
and with the methods of \cite{Feulner}
are described in this paper.


\section{The general setup.} \label{allg}

Throughout this section we let $G\leq S_n$ be an abelian group of odd order. 

The main strategy to construct self-dual $G$-invariant codes 
 $C=C^{\perp } \leq \F_2^n$ 
is  a bijection between these codes and
tuples 
$$(C_0,C_1,\ldots ,C_r,C_{r+1},C_{r+2},\ldots ,C_{r+2s})$$ of 
linear codes over extension fields of $\F_2$ 
that satisfy $C_0 = C_0^{\perp }$, $C_i = \overline{C_i}^{\perp }$ ($1\leq i \leq r$) 
and $C_{r+2i} = C_{r+2i-1}^{\perp }$ ($1\leq i \leq s$) for suitable inner 
products (see Lemma \ref{dual}).
Lower bounds on the minimum weight of $C$ will give rise to lower bounds
on suitably defined weights for the codes $C_i$ (see Lemma \ref{weight}). 
This gives a method to enumerate $G$-invariant self-dual codes with
high minimum weight. 

To this aim we view the $G$-invariant codes $C \leq \F_2^n$ as 
$\F_2G$-submodules of the permutation module $\F_2^n$, where 
$\F_2G$ is the group algebra of $G$. 
By Maschke's theorem this is a commutative semisimple algebra and hence 
a direct sum of fields. The codes $C_i$ from above will arise as linear 
codes over these direct summands of $\F_2G$. 

The underlying theory is well known and we do not claim to prove anything new in this section.
However we try to be 
very explicit here and therefore restrict to the special 
case that is relevant for the computations described in this paper. 
For the basic facts about representation theory of finite groups we refer the 
reader to \cite[Chapter VII]{HuB} and \cite[Chapter V]{Hup}.

\subsection{Abelian semisimple group algebras.}

$G$-invariant codes in $\F_2^n$ are modules for the group algebra
$$\F_2G := \{ \sum _{g\in G } a_g g \mid a_g \in \F_2 \} .$$ 
By Maschke's theorem  \cite[Theorem V.2.7]{Hup}
the group algebra $\F_2G $ is a commutative
semisimple algebra, i.e. a direct sum of fields. More precisely
$$\F_2 G \cong  \F_2 \oplus \F_{2^{k_1}} \oplus  \ldots \oplus \F _{2^{k_t}} $$
with $|G| = \dim _{\F_2}(\F_2 G) = 1 + k_1 + \ldots  + k_t $ and $k_i \geq 2$ for $i=1,\ldots , t$.
The projections $e_0,e_1,\ldots , e_t$ onto the simple components of $\F_2G$ (the central
primitive idempotents of $\F_2G$) can be computed as explicit linear
combinations of the group elements.
For instance $e_0 = \sum _{g\in G} g $,
expressing the fact that the first summand corresponds to the trivial representation
in which all group elements act as the identity.
In general any $g\in G$ defines an element $$ge_i \in \F_2 G e_i \cong \F_{2^{k_i}}  $$
of the extension field $\F_{2^{k_i}}$ of $\F_2$ 
and then $e_i = \sum _{g\in G} a_g g $ where $a_g = \trace_{\F_{2^{k_i}}/ \F_2 } (g^{-1} e_i)  $.

\begin{example} \label{z3z3} {\rm
Let
$G =\langle g,h \rangle \cong Z_3 \times Z_3$.
Since already $\F_4$ contains an element of order 3 $$\F_2G \cong
\F_2 \oplus \F_4 \oplus \F_4 \oplus \F_4 \oplus \F_4  .$$
If $h$ acts as the identity on $\F_2Ge_1 \cong \F_4$ and $g$ as a primitive
third root of unity, then the trace of $g^i h^je_1$ is $1$ if $i=1,2$
and $0$ if $i=0$. So $e_1 = (1+h+h^2)(g+g^2) $.
The coefficients of all the idempotents $e_i$  are given in the following table:
$$ \begin{array}{l|ccc|ccc|ccc|}
 & 1 & g & g^2 & h & gh & g^2h & h^2& gh^2 & g^2h^2 \\
\hline
e_0 & 1 & 1 & 1 & 1 & 1 & 1 & 1 & 1 & 1 \\
e_1 & 0 & 1 & 1 & 0 & 1 & 1 & 0 & 1 & 1 \\
e_2 & 0 & 0 & 0 & 1 & 1 & 1 & 1 & 1 & 1 \\
e_3 & 0 & 1 & 1 & 1 & 0 & 1 & 1 & 1 & 0 \\
e_4 & 0 & 1 & 1 & 1 & 1 & 0 & 1 & 0 & 1 \\
\hline \end{array}
$$
 }
\end{example}

The group algebra $\F_2G$ 
always carries a natural involution $$\overline{\phantom{a}} : \F_2 G \to \F_2 G, \sum _{g\in G} a_g g \mapsto 
 \sum _{g\in G} a_g g ^{-1} .$$ 
If $|G| > 1$ then this is an algebra automorphism of order 2.
It permutes the central primitive idempotents $\{ e_0,\ldots , e_t \}$. 
We always have $\overline{e_0} = e_0$ and order the idempotents such that  
$$\begin{array}{lll} \overline{e_i} = e_i & \mbox{ for } & i=0,\ldots ,r \leq t \\
\overline{e_{r+2i-1}} = e_{r+2i} & \mbox{ for } & i=1,\ldots , s
\end{array} $$
where $t=r+2s$. 

For later use we need explicit isomorphisms 
$$\tilde{\varphi }_i : \F_{2^{k_i}} \to \F_2 G e_i $$ 
that are compatible with the involution $\overline{\phantom{a}}$.
For $i=0$ there is just one 
$$\tilde{\varphi }_0 : \F_2 \to \F_2 G e_0 , 0\mapsto 0, 1 \mapsto e_0 .$$

\begin{lemma} \label{varphitilde}
\begin{itemize}
\item[(a)]
If $i\geq 1$ and $e_i = \overline{e_i} $ then $k_i$ is even and there is a 
unique automorphism $\sigma \in \Aut (\F_{2^{k_i}}) $ of order 2.
Then 
$$\tilde{\varphi }_i(\sigma (a)) = \overline{\tilde{\varphi}_i(a)} $$ 
for any isomorphism $\tilde{\varphi }_i$ and all $a\in \F_{2^{k_i}}$.
\item[(b)]
If $e_i \neq \overline{e_i} = e_j $, then $k_i = k_j$ and we may and will define
the pair $(\tilde{\varphi }_i,\tilde{\varphi }_j) $ such that  $\tilde{\varphi _j} = \overline{\tilde{\varphi _i}}$ so
$$\tilde{\varphi }_j: \F_{2^{k_j}} \to \F_2 G e_j , \tilde{\varphi }_j (a) = \overline{ \tilde{\varphi }_i (a)} $$
for all $a\in \F_{2^{k_j}}$.
\end{itemize}
\end{lemma}

\bew
(a) The fact that $k_i$ is even is a special case of Fong's theorem (see \cite[Theorem VII.8.13]{HuB}). 
In particular there is a
unique automorphism $\sigma \in \Aut (\F_{2^{k_i}}) $ of order 2. 
Since $a \mapsto \tilde{\varphi }_i ^{-1} (\overline{\tilde{\varphi}_i(a)} ) $ is an automorphism of 
$\F_{2^{k_i}}$ of order 1 or 2, we only need to show that this automorphism is not
the identity. 
Since  $\{ \tilde{\varphi }_i^{-1}(ge_i) \mid g\in G \}$ generates
 $\F_{2^{k_i}}$ over $\F_2$ and $k_i\geq 2$, there is some
$g\in G$ such that $ge_i \neq e_i$. 
Then $1\neq \tilde{\varphi}_i^{-1} (ge_i) =:a \in \F_{2^{k_i}}^*$ 
is a non-trivial invertible element and hence has odd order.
In particular $a\neq a^{-1}$ and so
$$\tilde{\varphi }_i ^{-1} (\overline{\tilde{\varphi}_i(a)} ) = 
\tilde{\varphi }_i^{-1}( g^{-1} e_i ) = a^{-1} \neq a .$$ 
(b) Clearly $k_i = k_j$ since under the assumption $\overline{\phantom{a}} : \F_2Ge_i \to \F_2G e_j $ 
is an isomorphism. The rest is obvious. 
\eb

\subsection{Invariant codes} \label{invaricodes}

To study all self-dual codes $C\leq \F_2^n$ such that
$ G \leq \Aut(C) $, 
we view $\F_2^n$ as an $\F_2G$-module where the elements $g\in G$
act by right multiplication with the corresponding permutation matrix $P_g\in \F_2^{n\times n}$.
So $\sum _{g\in G} a_g g  \in \F_2 G$ acts as $\sum _{g\in G} a_g P_g \in \F_2^{n\times n}$.
This way one obtains matrices $E_i \in \F_{2}^{n\times n }$ for the action of the idempotents $e_i \in \F_2G$,
where $E_iE_j = \delta _{ij}E_i$ and $E_0 + \ldots + E_t = 1$.
Then $\F_2^n$ is the direct sum
$$\F_2 ^n = \bigoplus _{i=0}^t \F_2^n E_i  .$$
The subspace $\F_2^n E_i$ is spanned by the rows of $E_i$. 
It is an
$\F_{2}Ge_i $-module,  hence a vector space over the finite field
$\F_{2^{k_i}} $.
So we may choose $\ell _i $ rows of $E_i$, say $(v_1,\ldots , v_{\ell _i})$,
to form an $\F_{2^{k_i}} $-basis of $\F_2^nE_i$.
We therewith obtain a non canonical isomorphism
\begin{equation} \label{deffi}
\varphi_i:  \F_{2^{k_i}} ^{\ell _i} \cong \F_2^n E_i, \varphi_i(a_1,\ldots , a_{\ell _i}) = \sum _{j=1}^{\ell _i} v_j \tilde{\varphi } _i(a_j) 
\end{equation}
for $i=0,\ldots , t$, where the isomorphisms $\tilde{\varphi _i}$ are as in Lemma \ref{varphitilde}.

Any $G$-invariant code $C$, being an $\F_2G$-submodule of $\F_2^n$, decomposes
uniquely as $$ C
= \bigoplus _{i=0}^t C E_i
= \bigoplus _{i=0}^t \varphi_i (C _i)  $$ for $\F_{2^{k_i}} $-linear codes $$C_i \leq  \F_{2^{k_i}}^{\ell _i} $$

\begin{lemma} \label{codes}
The mapping 
$$\varphi :  (C_0,C_1,\ldots , C_t) \mapsto \bigoplus _{i=0}^t \varphi_i (C _i)  $$ 
is a bijection between the set 
$${\mathcal C}_G := \{ (C_0,C_1,\ldots , C_t) \mid C_i \leq \F_{2^{k_i}}^{\ell _i } \} $$ and 
the set of $G$-invariant codes in $\F_2^{n}$.
\end{lemma} 

So instead of enumerating directly the $G$-invariant codes $C\leq \F_2^n$ 
we may enumerate all $(t+1)$-tuples of linear codes $C_i \leq \F_{2^{k_i}}^{\ell _i } $.
Comparing the $\F_2$-dimension we get $n=\sum _{i=0}^{t} k_i \ell _i $, 
so the length $\ell _i$ is usually much smaller than $n$. 

\subsection{Duality} 

We are interested in self-dual codes with respect to the 
 standard inner product 
$$v\cdot w := \sum _{i=1}^n v_i w_i $$ on $\F_2^n$. 
This is invariant under permutations, so
$ vg \cdot wg = v\cdot w  $ for all $v,w\in \F_2^n$ and $g\in S_n$.
We hence obtain the equation
\begin{equation}\label{eq:adjoint}
 vg \cdot w = v \cdot wg^{-1} \mbox{ for all } v,w\in \F_2^n, g\in S_n .
\end{equation}
This tells us that the adjoint of a permutation $g$ with respect to the inner product is $\overline{g} = g^{-1}$,
for the natural involution $\overline{\phantom{a}} $ of $\F_2G$.
From Equation \eqref{eq:adjoint} we  hence obtain that
$$ va \cdot w = v \cdot w \overline{a} \mbox{ for all } v,w\in \F_2^n, a\in \F_2 G .$$
In particular the idempotents of $\F_2G$ satisfy 
\begin{equation}\label{eq:orth1}
vE_i \cdot w E_j = v \cdot w E_j \overline{E_i} \mbox{ for all } v,w\in \F_2^n .
\end{equation}
Since $E_j \overline{E_i} = 0$ if $E_i\neq \overline{E_j}$ we hence obtain an 
orthogonal decomposition 
\begin{equation}\label{eq:orth}
\begin{array}{ll}
 \F_2^n  & = \perp _{i=0}^r \F_2^n E_i \perp \perp _{j=1}^s (\F_2^n E_{r+2j-1} \oplus \F_2^n E_{r+2j} ) = \\
& 
  \perp _{i=0}^r \F_2^n E_i \perp \perp _{j=1}^s (\F_2^n \overline{E}_{r+2j} \oplus \F_2^n E_{r+2j} ) 
\end{array} 
\end{equation}

\begin{definition} \label{defsp} 
For $0\leq i \leq t$ let $\varphi _i : \F_{2^{k_i}} ^{\ell _i} \to \F_2^n E_i $ be the
isomorphism from Equation \eqref{deffi}. 
For $0\leq i \leq r $ define the inner product 
$$h_i : \F_{2^{k_i}}^{\ell _i} \times \F_{2^{k_i}}^{\ell _i} \to \F_2 , h_i(c,c') := \varphi _i(c) \cdot \varphi_i(c') $$ 
and use $h_i$ to define the dual of a code $C_i \leq \F_{2^{k_i}}^{\ell _i}$ as 
$$C_i^{\perp } := \{ v\in \F_{2^{k_i}}^{\ell _i} \mid h_i(v,c) = 0 \mbox{ for all } c\in C_i \} .$$
For $j=1\ldots , s$ let $J:=r+2j$ and define 
$$s_j : \F_{2^{k_{J}}}^{\ell _{J}} \times \F_{2^{k_{J-1}}}^{\ell _{J-1}} \to \F_2 , s_j(c,c') := 
\varphi _{J}(c) \cdot \varphi_{J-1}(c') .$$
Then $s_j$ defines the dual $C_{J-1}^{\perp } \leq \F_{2^{k_J}}^{\ell _J} $ of a code $C_{J-1} \leq \F_{2^{k_{J-1}}}^{\ell _{J-1}} $
as 
$$ C_{J-1}^{\perp } := \{ v\in \F_{2^{k_J}}^{\ell _J} \mid s_j (v,c) = 0 \mbox{ for all } c\in C_{J-1} \} .$$
\end{definition}

\begin{lemma} \label{dual}
Let $C=\varphi (C_0,\ldots , C_t)  \leq \F_2^n$ be some $G$-invariant code. Then the dual code is
$C^{\perp } = C'$ where 
$$ C' := \varphi (C_0^{\perp }, C_1^{\perp } , \ldots , C_r^{\perp }, 
C_{r+2}^{\perp}, C_{r+1}^{\perp },\ldots , C_{t}^{\perp }, C_{t-1}^{\perp } )  .$$
In particular the set of all self-dual $G$-invariant codes $C=C^{\perp } \leq \F_2^n$ 
is the image (under the bijection $\varphi $ of Lemma \ref{codes}) of the set 
$$\begin{array}{ll} 
{\mathcal C}_G^{sd}:= & \{ (C_0, C_1,\ldots , C_t)  \in {\mathcal C}_G \mid C_i=C_i^{\perp} (0\leq i \leq r) \\ & 
C_{r+2j} = C_{r+2j-1}^{\perp } (j=1,\ldots , (t-r)/2) \} .
\end{array} $$
\end{lemma} 

\bew
Comparing dimension it is enough to show that 
$C^{\perp } \supseteq  C' .$
Since $C = \bigoplus _{i=0}^t C E_i$ and 
$$C' = \bigoplus _{j=0}^r \varphi _j(C_j^{\perp}) \oplus \bigoplus _{j=1}^{s} \varphi _{r+2j-1} (C_{r+2j}^{\perp} ) 
\oplus \varphi _{r+2j} (C_{r+2j-1}^{\perp} ) $$ it suffices to show that any element of $CE_i$ is 
orthogonal to any component of $C'$. 
\\
So let $c\in C_i$ and 
first assume that $i\leq r$. 
By Equation \eqref{eq:orth1}  $$\varphi _i(c) \cdot \varphi _j(c') = 0 \mbox{ for all } j\neq i \mbox{ and } c'\in  \F_{2^{k_j}}^{\ell } .$$
For $j=i$ we compute 
$$ \varphi _i(c) \cdot \varphi_i(c') = h_i(c,c') \mbox{ for all } 
 c'\in \F_{2^{k_i}}^{\ell } .$$
This is $0$ if $c'\in C_i^{\perp }$.
\\
Now assume that  $i = r+2k$.  
Then Equation \eqref{eq:orth1} yields  $$\varphi _i(c) \cdot \varphi _j(c') = 0 \mbox{ for all } j\neq r+2k-1 \mbox{ and } c'\in  \F_{2^{k_j}}^{\ell } .$$
For $j=r+2k-1$ we have 
$$\varphi _{r+2k}(c) \cdot \varphi _{r+2k-1} (c') = s_k (c,c') \mbox{ for all } c' \in \F_{2^{k_j}}^{\ell }.$$
This is $0$  if $c' \in C_{r+2k}^{\perp } $. 
\\
A similar argument holds for $i=r+2k-1$.
\eb

\subsection{Weight} 

Enumerate the group elements so that 
 $G= \{ 1=g_1,\ldots , g_q \} \leq S_n$ with $q = |G|$. Then by assumption $q$ is odd.

\begin{lemma} \label{block}
Assume that $G\leq S_n$ fixes the points $m+1,\ldots , n $ and that every element 
$1\neq g\in G$ acts without any fixed points on $\{ 1,\ldots , m\} $.
Then $$\ell _i = \ell  = \frac{m}{q} $$ for all $i>0$ and 
after reordering the elements in $\{ 1,\ldots , m\}$ and therewith replacing 
$G$ by a conjugate group we may assume that 
$$g_i(kq+1) = kq+i $$ 
for all $i=1,\ldots , q$, $k=0,\ldots , \ell -1 $.
\end{lemma} 

\bew
For $j\in \{ 1,\ldots ,m \}$ the stabiliser in $G$ of $j$ consists only of the identity 
and hence the orbit $Gj = \{ g_1(j),\ldots , g_q(j) \}$ has length $q$ and therefore 
$m=\ell q$ is a multiple of the group order $q=|G|$. 
From each of the $\ell $ orbits choose some element $j_k$.  
The reordering is now obviously 
$$(g_1(j_1), g_2(j_1) ,\ldots , g_q(j_1), g_1(j_2),
\ldots , g_q(j_{\ell }) ) .$$
\eb

In this new group the permutation matrices $P_g$ are block diagonal matrices with 
$\ell $ equal blocks of size $q$ and an identity matrix $I_{n-m}$ of size $n-m$ at the
lower right corner. 
Also the idempotent matrices $E_i$
 are block diagonal 
$$\begin{array}{ll} 
E_0 = \diag (B_0,\ldots , B_0 , I_{n-m})  \\
 E_i = \diag (B_i,\ldots , B_i , 0_{n-m} )  & 
1\leq i \leq t . \end{array} $$
If $e_i = \sum _{k=1}^q \alpha _k g_k $, then the first row of 
$B_i$ is $(\alpha _1,\ldots , \alpha _q)$ and the other rows of $B_i$ are obtained by 
suitably permuting these entries.
The rank of the matrix $B_i$ is exactly $k_i$. Let
$$\eta _i: \F_2G e_i \to \mbox{rowspace}(B_i), \sum _{k=1}^q \epsilon _k g_k e_i \mapsto  (\epsilon _1,\ldots,\epsilon _q) B_i .$$
Then the isomorphism 
 $\varphi _i: \F_{2^{k_i}} ^{\ell } \to \F_2^n E_i  \leq \F_2^n $ 
is defined by 
$$\varphi _i (c_1,\ldots , c_{\ell }) := (\eta_i(\tilde{\varphi }_i(c_1)) , 
\eta _i (\tilde{\varphi }_i(c_2)), \ldots , \eta_i(\tilde{\varphi }_i(c_{\ell })) ) .$$

\begin{lemma} \label{weight}
In the situation above  define 
 a weight function $w_i : \F_{2^{k_i}} \to \Z _{\geq 0}$ 
by 
$$w_i (x) := \wt (\eta _i (\tilde{\varphi _i} (x)) ) .$$
If $i\geq 1$ or $m=n$, 
then 
$$\wt_i : \F_{2^{k_i}} ^{\ell }  \to \Z _{\geq 0}, c\mapsto \sum _{k=1}^{\ell } w_i(c_k) $$
defines a weight function on $\F_{2^{k_i}}^{\ell }$ 
such that the isomorphism $\varphi_i$
is weight preserving.
\end{lemma}

\bew
We need to show that 
$\wt (\varphi _i(c)) = \wt _i (c) $ for all $c\in \F_{2^{k_i}}^{\ell } $. 
But 
$$\varphi _i ((c_1,\ldots , c_{\ell }) ) 
= (\eta_i(\tilde{\varphi }_i(c_1)) , 
\eta _i (\tilde{\varphi }_i(c_2)), \ldots , \eta_i(\tilde{\varphi }_i(c_{\ell })),0^{n-m} ) $$
and so the weight of $\varphi _i(c )$ is the sum 
$$\wt (\varphi _i(c)) = \sum _{k=1}^{\ell } \wt (\eta_i(\tilde{\varphi }_i(c_k) ) ) 
= \sum _{k=1}^{\ell } w_i(c_k) .$$
\eb

\begin{rem} \label{weightspec0} {\rm 
For $m<n$ and $i=0$, we need to modify the weight function because we work 
with $\ell $ blocks of size $q$ and $n-m$ blocks of size 1. 
So here 
$\wt_0 : \F_2^{\ell + (n-m)} \to \Z_{\geq 0}  $  $$
\begin{array}{l} 
\wt _0 (c_1,\ldots , c_{\ell } , d_1,\ldots, d_{n-m}) = \\ q\wt (c_1,\ldots , c_{\ell }) 
+ \wt (d_1,\ldots , d_{n-m }) . \end{array} $$
}
\end{rem} 

\begin{rem} \label{weightspecequiv} {\rm 
We will always work with $G$-equivalence classes of codes, 
where $C, C'\leq \F_2^n$ are called {\em $G$-equivalent}, if there is 
some permutation 
$$\pi \in S_{n,G}:= \{ \pi \in S_n \mid \pi g = g \pi  \mbox{ for all } g\in G \} $$
mapping $C$ to $C'$. 
In the situation of Lemma \ref{block} the group 
$$S_{n,G}  \cong G \wr S_{\ell } \times S_{n-m} $$
is obtained by the action of $G$ on the blocks of size $q$ and the 
symmetric group $S_{\ell }$ permuting the $\ell $ blocks of size $q$.
The group $S_{n-m}$ permutes the last $n-m$ entries. 
Via the isomorphism  $\varphi _i$  constructed in Lemma \ref{weight} the 
action of $S_{n,G}$ on $\F_2^n E_i \cong \F_{2^{k_i}}^{\ell }$ translates into
the monomial action with monomial entries in the subgroup 
$$\langle \varphi _i^{-1} (ge_i) \mid g\in G \rangle \leq \F_{2^{k_i}}^* .$$
Note that these are weight preserving automorphisms of the space 
$ \F_{2^{k_i}} ^{\ell }$ for the weight function defined in Lemma \ref{weight}.
}
\end{rem}

\begin{rem} {\rm
\label{weightspecprod} 
For the weight preserving isomorphisms $\varphi _i$ constructed in Lemma \ref{weight} 
the inner product $h_i$ and $s_j$ defined in Definition \ref{defsp} are
standard inner products:  For $0\leq i \leq r$  and $c,c' \in \F_{2^{k_i}}^{\ell } $ 
$$ h_i(c,c' ) = 
\sum _{k=1}^{\ell } \eta_i (\tilde{\varphi }_i (c_k)) \cdot \eta_i( \tilde{\varphi }_i (c'_k))  .$$ 
For $1\leq j \leq s$ with $J:=r+2j$, $c\in \F _{2^{k_J}} ^{\ell }, c' \in \F _{2^{k_{J-1}}} ^{\ell }$
$$ s_j (c,c') = 
 \sum _{k=1}^{\ell } \eta_J( \tilde{\varphi }_{J} (c_k)) \cdot \eta_{J-1}(\tilde{\varphi }_{J-1} (c'_k))  $$
}
\end{rem}


\subsection{Strategy of computation.} 

Now the computational strategy to enumerate representatives of 
the $G$-equivalence classes of all self-dual $G$-invariant codes 
$C= C^{\perp } \leq \F_2^n$ with minimum weight $d$ is as follows: 
We successively enumerate the codes $C_0$, $C_1$, $\ldots $ such that 
$(C_0,\ldots , C_t ) \in {\mathcal C}_G^{sd} $ yields a self-dual $G$-invariant code 
by Lemma \ref{dual}. 
With Lemma \ref{weight} we control the minimum weight of $\varphi _i(C_i)$
using the suitable weight function $\wt _i$ on $\F_{2^{k_i}}^{\ell }$. 
We only continue with those codes $ (C_0,\ldots , C_i) $ for which 
$$\bigoplus _{j=0}^i \varphi _j (C_j) \leq \F_2^n $$ has minimum weight  $\geq d$.
Equivalence translates into the monomial equivalence from Remark \ref{weightspecequiv}. 
We have a simultaneous action of the monomial group 
$${\mathcal M} := \langle (\varphi_0^{-1} (ge_0), \ldots , \varphi _t^{-1} (ge_t)) \mid g\in G \rangle \wr S_{\ell } \times S_{n-m}  .$$
If we have already found the tuple $(C_0,\ldots , C_i) $ then 
only the stabiliser in ${\mathcal M}$ of these $i+1$ codes acts 
on the set of candidates for $C_{i+1}$.

\section{The case $Z_3\times Z_3$.} 

From now on 
let $C \leq \F_2^{72} $ be a binary self-dual code with minimum distance 16.
Then $C$ is doubly-even (see \cite{Rains}) and hence an extremal 
Type II code. 

In this section we assume that $\Aut (C)$ contains a subgroup $G$ isomorphic to
$Z_3 \times Z_3$. 
By \cite[Theorem 1.1]{XXX} any element of order 3 in $\Aut(C)$ acts without fixed points 
on $\{ 1,\ldots , 72 \}$, so $G$ is conjugate in $S_{72}$ to the subgroup 
 $ G=\langle g,h \rangle \leq S_{72}$
where 
$$\begin{array}{lcl} g &=& (1,4,7)(2,5,8)(3,6,9) \ldots (66,69,72) \\ 
h & = & (1,2,3)(4,5,6)(7,8,9)  \ldots (70,71,72) \end{array}  $$

The following Lemma gives the structure of the 
fixed code of any  $1\neq g\in G$.

\begin{lemma} (cf. \cite{Huffman}) 
Let $C$ be a Type II code of length $72$ and minimum distance $16$
and let $g\in \Aut(C)$ be an automorphism of order $3$.
Then the fixed code of $g$ in $C$ is equivalent to 
${\mathcal G}_{24} \otimes \langle (1,1,1) \rangle $, where 
${\mathcal G}_{24} \leq \F_2^{24}$ is the extended binary Golay code,
 the unique binary $[24,12,8]$-code.
\end{lemma}

\bew
We apply the methods of Section \ref{allg} to the group $\langle g \rangle \leq S_{72}$. 
Let $E_0:=1+P_g+P_g^2 \in \F_2^{72\times 72}$. 
Then $E_0$ is the projection onto the fixed space of $g$, 
$\F_2^{72} E_0 \cong \varphi_0(\F_2^{24}) $ 
and $CE _0 = \varphi _0 ({\mathcal G})$ for some self-dual  
 binary code ${\mathcal G} \leq \F_2^{24}$ (see Lemma \ref{dual}). 
Since $C$ is doubly-even, also ${\mathcal G}$ is a Type II code.
Moreover the minimum distance of $\varphi_0({\mathcal G} ) $ 
is 3 times the minimum distance of ${\mathcal G}$ (see Lemma \ref{weight}). 
Since $CE_0 \leq C$ has minimum
distance $\geq 16$, we conclude that the minimum distance of 
${\mathcal G}$ is $\geq 6$ and hence $\geq 8$ since ${\mathcal G}$ is doubly-even. 
This shows that ${\mathcal G}$ is equivalent to the Golay code.
\eb

\begin{rem} \label{D1} {\rm
Let 
$$C(h) := \{ c\in C \mid ch=c \} \cong {\mathcal G} \otimes \langle (1,1,1) \rangle $$ 
be the fixed code of $h$.
Then $g$ acts as an automorphism $g'$ on the Golay code ${\mathcal G}$ and has no 
fixed points on the places of ${\mathcal G}$. 
Up to conjugacy in $\Aut ({\mathcal G})$ there is a unique such automorphism $g'$. 
We use the notation of Section \ref{allg} for $G':=\langle g' \rangle \leq S_{24}$.
To distinguish the isomorphisms $\varphi _i$ from  those defined by $G$, 
we use the letter $\psi $ instead of $\varphi $. 
As an $\F_2\langle g'\rangle $ module the code ${\mathcal G}$ decomposes as 
$$ {\mathcal G}= \psi _0 (D_0 ) \perp \psi _1 (D_1) .$$ 
Explicit computations show that 
$D_0 \cong h_8 \leq \F_2^8$ is the extended 
Hamming code $h_8$ of length 8
and $D_1 \cong \F_4 \otimes _{\F_2 } h_8$.  }
\end{rem}

We now use the isomorphisms $\varphi _i$ constructed in Section \ref{invaricodes} 
for the group $G=\langle g,h \rangle \cong Z_3 \times Z_3$ and the 
idempotents $e_0,\ldots ,e_4$ from Example \ref{z3z3}. 
Since all the $e_i$ are invariant under the natural involution 
the extremal $G$-invariant code $C=C^{\perp } \leq \F_2^{72}$ decomposes as
$$C = \perp  \varphi_i (C_i ) $$ 
for some self-dual Type II code $C_0 \leq \F_2^{8}$ 
and Hermitian self-dual codes $C_i  \leq \F_4^8$. 
Then all the $C_i$ (for $i=1,2,3,4$) are equivalent to the 
code $D_1$ from Remark \ref{D1} and hence $C_i \cong \F_4 \otimes _{\F_2} h_8$ for all $i=1,2,3,4$.

\begin{rem} {\rm 
$C_i \cong \F_4 \otimes _{\F_2} h_8$ for all $i=1,2,3,4$. 
Moreover for all $i=1,2,3,4$ the code
$$ \psi _0(C_0 )  \oplus  \psi _i (C_i ) \cong {\mathcal G} $$
is equivalent to the binary Golay code of length $24$. 
}
\end{rem}

The main result of this section is the following theorem. 

\begin{theorem} \label{main1}
There is no extremal self-dual Type II code $C$ of length $72$ 
for which $\Aut(C)$ contains $ Z_3 \times Z_3$.
\end{theorem}

\bew
For a proof we describe the computations that led to this result using the
notation from above.
To obtain all candidates for the codes $C_i$ we first 
fix a copy $C_0 \leq \F_2^8$ of the Hamming code $h_8$. 
We then compute the 
orbit of $\F_4 \otimes _{\F_2} h_8$ under the full monomial group 
$\F_4^* \wr S_8$ and check for all these codes $C_i \leq \F_4^8$ whether 
$ \psi_0(C_0) \oplus \psi _1(C_i) $
has minimum distance $8$. 
This yields a list 
${\mathcal L}$ of $17,496$  candidates for the codes $C_i\leq \F_4^8$.

Since there is up to equivalence a unique Golay code and this code has
a unique conjugacy class of fixed-point free automorphisms $g'$ of order $3$, 
we may choose a fixed representative 
for $C_0\leq \F_2^8$ and $C_1\leq \F_4^8$. 
The centralizer of $g'$ in the automorphism group of
$$ {\mathcal G} = \psi _0 (C_0 )  \perp  
\psi _1 (C_1 ) $$ 
acts on ${\mathcal L}$ with 138 orbits. 
Choosing representatives $C_2$ of these orbits, we obtain 138 
doubly even binary codes
$$ D = \varphi _0(C_0) \oplus \varphi _1(C_1) \oplus \varphi_2(C_2) $$
of length 72, dimension $20$, and minimum distance $\geq 16$. 
These codes $D$ fall into 2 equivalence classes under the action of the
full symmetric group $S_{72}$. 
The automorphism group of both codes $D$ contains up to 
conjugacy a unique subgroup $U$
$\cong Z_3\times Z_3$ that has 8 orbits of length 9 on $\{1,\ldots , 72\}$ 
and such that there are generators $g,h$ of $U$ each having a 12-dimensional 
fixed space on $D$. 
For both codes $D$ we compute the list 
$${\mathcal L}_3(D) := \{ C_3 \in {\mathcal L} \mid 
d(D \oplus \varphi_3(C_3)  ) \geq 16 \} $$
and similarly 
$${\mathcal L}_4(D)   := 
 \{ C_4 \in {\mathcal L} \mid 
d(D \oplus \varphi_4(C_4)  ) \geq 16 \} .$$
The cardinalities are 
$$|{\mathcal L}_3(D)| = |{\mathcal L}_4(D) | =  7146 \mbox{ or } 2940 .$$
It takes about 2 days of computing time to go through the list of 
pairs $(C_3,C_4) \in {\mathcal L}_3(D)\times {\mathcal L}_4(D) $ and check 
whether 
$D \oplus \varphi_3(C_3) \oplus \varphi_4(C_4) $
  has minimum distance $\geq 16$ using Magma \cite{MAGMA}.
No extremal code is found.
\eb

\section{Automorphisms of order seven.} \label{sec:Order_seven}

Let $C=C^{\perp } \leq \F_2^{72}$ be an extremal Type II code.
Assume that there is an element $g\in \Aut (C) $ of order 7. 
Then by \cite[Theorem 6]{ConwayPless} the permutation $g\in S_{72}$ is the 
product of 10 seven-cycles.
Wlog we assume that 
$$g = (1,\ldots , 7) (8,\ldots, 14) \cdots (83,\ldots, 70) $$
fixes the points $71$ and $72$, so in the notation of Lemma \ref{weight} $m=70$. 
The central primitive idempotents 
$$e_0 = \sum _{i=0}^6 g^i ,\ e_1 = g^4+g^2+g+1 , \ e_2 = g^6+g^5+g^3+1 $$
of $\F_2 \langle g \rangle$ satisfy 
$$\overline{e_1} = e_2 \mbox{ and } \F_2 \langle g \rangle e_i \cong \F_8  \mbox{ for } i=1,2. $$
In the notation of Section \ref{allg} the code $C$ is of the form
$$ C = \varphi _0 (C_0) \perp \varphi _1 (C_1) \oplus \varphi _2(C_1^{\perp } ) $$ 
for some self-dual code 
$$C_0 = C_0 ^{\perp } \leq \F_{2}^{10+2} $$ 
and $C_1  \leq \F_8^{10} $. 
To obtain weight preserving isomorphisms $\varphi _i $ we consider the kernel $D$ of
the projection of $C$ onto the last 2 coordinates.  
So let $$D_0 := \{ (c_1,\ldots , c_{10} ) \mid (c_1,\ldots , c_{10},0,0) \in C_0 \} $$
and define $D:=\varphi _0(D_0) \perp \varphi _1 (C_1) \oplus \varphi _2(C_1^{\perp }) \leq \F_2^{70 }$
Then 
$$D = \{ (c_1,\ldots , c_{70}) \mid (c_1,\ldots , c_{70},0,0) \in C\}  $$ is a  doubly-even code 
of dimension 34 and minimum distance $\geq 16$. 
Applying Lemma \ref{weight} and Lemma \ref{dual} to this situation one finds the conditions
$$\begin{array}{ll} D_0 \subset D_0^{\perp } \leq \F_2^{10} & \mbox{ doubly even } \\
C_1 \leq \F_8^{10}  & d(C_1) \geq 4 , d(C_1^{\perp }) \geq 4 . \end{array} $$
We hence compute the linear codes $C_1\leq \F_8^{10}$ such that 
$d:=d(C_1) \geq 4$ and the dual distance $d^{\perp } = d(C_1^{\perp })  \geq 4$. 
For each such code $C_1$ we check if the code 
$$\tilde{C}_1 := \varphi _1 (C_1) \oplus \varphi _2(C_1^{\perp}) \leq \F_2^{70 } $$ 
has minimum distance $\geq 16$.

\begin{table}[bt]
\begin{center}
\begin{tabular}{ccc|r|r}
\multicolumn{3}{c|}{Parameters of $C_1$} & \multicolumn{2}{c}{Number of 
non isomorphic candidates }\\
$k$ & $d$ & $d^{\perp}$ &  for $C_1$ & for $C_1$ with $d( \tilde{C}_1  ) 
\geq 16$
\\\hline
3 & 8 & 4 & 1 & 1\\
4 & 4 & 4 & 81,717 & 657 \\
4 & 5 & 4 & 1,854,753 & 8,657\\
4 & 6 & 4 & 490,382 & 2,632\\
5 & 4 & 4 & 61,487,808 & 145,918 \\
5 & 5 & 4 & 3,742,898 & 10,769 \\ 
5 & 5 & 5 & 3,014,997 & 9,216\\ \hline
\multicolumn{3}{c|}{Total} & 70,672,556 & 177,850
\end{tabular}
\end{center}
\caption{Computational Results for $Z_7$}
\label{table:classes}
\end{table}

\begin{lemma}\label{E}
If $C$ is an extremal Type II code then 
$D_0$ is equivalent to 
the maximal doubly-even subcode $E$ 
of the 2-fold repetition code $\F_2^5 \otimes \langle(1,1)\rangle$.
\end{lemma}

\bew
Clearly $D_0\leq \F_2^{10}$ is doubly-even and of dimension $4$, 
$$D_0^{\perp}  >  A_0, A_1,A_2 > D_0 $$ with $A_0 = A_0 ^{\perp }$ a Type I code 
and $A_2 = A_1^{\perp }$. The code $C_0$ is a full glue of 
$D_0^{\perp }/D_0$ and $\F_2^2$, 
$$ \begin{array}{l} C_0  =   \{ (a,1,1) \mid a\in A_0 \setminus D_0 \} 
 \disj  
 \{ (a,0,0) \mid a\in  D_0 \}  \\
 \disj \{ (a,1,0) \mid a\in A_1 \setminus D_0 \}   \disj  
 \{ (a,0,1) \mid a\in A_2 \setminus D_0 \}  
\end{array} $$
For $a\in D_0^{\perp }$ and $x\in \F_2^2$ the weight
$$\wt (\varphi_0(a,x) ) = 7 \wt (a) + \wt (x) $$
because $\varphi _0$ repeats the first 10 coordinates 7 times (see Remark \ref{weightspec0})
 and leaves the last two unchanged. 
Since $\varphi_0(C_0)$ has minimum distance $\geq 16$, the set $A_1\cup A_2 $ needs to have minimum
weight $>2$. Since the weights of the words in the set $D_0^{\perp } \setminus A_0$, 
the shadow of $A_0$ in the sense of \cite[p. 1320]{CS},
 are $\equiv \frac{10}{2} \pmod{4}$, the minimum weight there 
needs to be 5. This forces $A_0 $ to be equivalent to 
$\F_2^5 \otimes \langle(1,1)\rangle$.
\eb

\begin{theorem} \label{main2}
There is no extremal self-dual Type II code of length $72$ that has an
automorphism of order $7$.
\end{theorem}

\bew
Based on the description of the code $D$ of length 70 above we use a 
computer search to show that no such code $D$ has minimum distance $\geq 16$. 
For this purpose we classify all codes in $C_1 \leq \F_8^{10}$ such that 
$C_1$ and its dual $C_1^{\perp } $ both have minimum distance $\geq 4$, 
see \cite{Feulner} for more details. Furthermore, it is sufficient to consider only one of the two
dual parameter sets $[10,k,d, {d^\perp}]$ and $[10,10-k,d^\perp, d]$ since the interchange of $C_1$ and $C_1^\perp$ leads to isomorphic codes. 

The maximal dimension of such a code $C_1$ is 7.
Up to semi-linear isometry there are more than 70 million such codes.
The condition that the minimum distance of the code
$\tilde{C}_1 := 
 \varphi_1( C_1 ) \oplus \varphi _2 (C_1^{\perp }) $ 
is $\geq 16$ reduces the number of codes to about 180,000 codes that need to 
be tested, see Table \ref{table:classes} for details.
For each of these codes $\tilde{C}_1$ we run through all 
945 different binary codes $D_0 \leq \F_2^{10}$ that are equivalent to $E$ 
from Lemma \ref{E} and check whether the code 
$D:= \varphi_0(D_0)  \oplus \tilde{C}_1 $
has minimum distance $\geq 16$.
No such code is found.
\eb

\section{The dihedral group of order 10} 

\subsection{Automorphisms of order 5.} 

Let $C=C^{\perp } \leq \F_2^{72}$ be an extremal Type II code.
Assume that there is some element $g\in \Aut (C) $ of order 5. 
Then by \cite[Theorem 6]{ConwayPless} the permutation $g\in S_{72}$ is the 
product of 14 five-cycles and we assume that 
$$ g = (1,2,3,4,5)(6,7,8,9,10) \ldots (66,67,68,69,70)  $$
The primitive idempotents in $\F_2 \langle g \rangle $ are
$$e_0 = \sum _{i=0}^4 g^i ,\ e_1 = 1+e_0 = g+g^2+g^3+g^4 $$
and $\F_2 \langle g \rangle e_1 \cong \F_{16} $.
As an $\F _2 \langle g \rangle $ submodule of $\F_2^{72}$,  the code $C$ decomposes as 
$$\varphi _0(C_0) \perp \varphi_1( C_1) ,\mbox{ with } C_0 =C_0^{\perp } \leq \F_2^{16}, C_1   = C_1^{\perp } 
\leq \F_{16}^{14 }.$$ 
As above let 
$D := \{ (c_1,\ldots ,c_{70} ) \mid (c_1,\ldots, c_{70},0,0) \in C \}  $. 
Then $D$ is a  doubly-even code in $\F_2^{70} $
of dimension 34 and minimum distance $\geq 16$ and 
$$D= \varphi _0(D_0) \perp \varphi _1(C_1) $$
for some doubly-even code $D_0 \leq \F_2^{14}$ of dimension 4. 

\begin{lemma}\label{E5}
If $C$ is an extremal Type II code then 
$D_0$ is equivalent to 
the maximal doubly-even subcode $E$ 
of the unique self-dual code $A_0 \leq \F_2^{14}$ of minimum distance $4$.
\end{lemma}

\bew
Clearly $D_0\leq \F_2^{14}$ is doubly-even and of dimension $6$, 
$$D_0^{\perp}  >  A_0, A_1,A_2 > D_0 $$ with $A_0 = A_0 ^{\perp }$ a Type I code 
and $A_2 = A_1^{\perp }$. 
As in the proof of Lemma \ref{E}, 
 code $C_0$ is a full glue of $D_0^{\perp}/D_0$ and $\F_2^2$.
For $a\in D_0^{\perp} $ and $x\in \F_2^2$ the weight of 
$$\varphi _0(a,x) \in \varphi _0(C_0) \leq C$$
is $5 \wt (a) + \wt (x) $. 
Since $\varphi _0(C_0)$ has minimum distance $\geq 16$, the code $A_0$ 
needs to have minimum weight $\geq 4$.
Explicit computations show that there is up to equivalence a unique such code $A_0$.
\eb

To obtain a weight preserving isomorphism $\varphi _1 : \F_{16}^{14} \to \F_2^{72} E_1 $
as described in Lemma \ref{weight} 
we need to define the suitable weight function on the coordinates $c_k \in \F_{16}$:

\begin{definition}
Let $\xi \in \F_{16}^*$ denote a primitive 5th root of unity.
The {\em 5-weight} of $x\in \F_{16}$ is 
$$\wtfive(x):= 
  \left\{ \begin{array}{cc} 0 & x = 0 \\ 
4 & x\in \langle \xi  \rangle \leq \F_{16}^* \\ 
2 & x \in \F_{16}^* \setminus \langle \xi  \rangle \end{array} \right. $$
For $c=(c_1,\ldots ,c_n) \in \F_{16}^n$ we let as usual
$\wtfive(c):= \sum _{i=1}^n \wtfive(c_i) $.
\end{definition}

\subsection{The dihedral group of order 10.} 

We now assume that $C=C^{\perp } \leq \F_2^{72}$ is an extremal Type II code
such that 
$$D_{10} \cong G := \langle g,h \rangle \leq \Aut (C) $$
where $g$ is the element of order 5 from above and the order of $h$ is 2.
By \cite{StefkaB} any automorphism of order 2 of $C$ acts without fixed points, so 
we may assume wlog that 
$$\begin{array}{l} g = (1,2,3,4,5)(6,7,8,9,10) \ldots (66,67,68,69,70) , \\
h = (1,6)(2,10)(3,9)(4,8)(5,7) \ldots \\ (61,66)(62,70)(63,69)(64,68)(65,67) \cdot (71,72) .
\end{array} $$
The centralizer in $S_{72}$ of $G$ isomorphic to 
$D_{10} \wr S_7\times \langle (71,72) \rangle $ acts on the set of $G$-invariant codes. 

\begin{rem} {\rm 
Let $e_0$ and $e_1=1+e_0\in \F_2 \langle g \rangle \leq \F_2 G $ be as above. 
Then $e_0$ and $e_1$ are the central primitive idempotents in $\F_2 G$. 
In particular $\langle h \rangle $ acts on the codes $CE_0$ and 
$CE_1$. }
\end{rem}

\begin{rem} \label{listeC0} {\rm 
Explicit computations with MAGMA show that 
the automorphism group of the code $A_0$ from Lemma \ref{E5} 
contains a unique conjugacy class of elements
$x$ of order 2 that have 7 orbits. 
Therefore the action of  $h$  on the fixed code of $\langle g \rangle$ is uniquely determined.
Let $U$ be
the centralizer of $x$ in the full symmetric group of degree 14.
Then the $U$-orbit $O_{14}$ of $A_0$ has length 1920. 
Let $$\mathcal{C}_0 := \{ \varphi _0(C_0) \mid C_0 \in O_{14} \} . $$
}
\end{rem}

To investigate the action of the element $h$ on the Hermitian self-dual code 
$C_1 \leq \F_{16}^{14} $  we recall the 
following theorem.

\begin{theorem} (\cite[Theorem 3.1]{NebeF2}) 
The fixed code $C(h):= \{ c\in C \mid ch=c \} $ of $\langle h \rangle $ is 
equivalent to $B\otimes \langle (1,1)\rangle $ for some self-dual code
$B=B^{\perp } \leq \F_2^{36}$ of minimum distance $8$. 
\end{theorem} 

Let $$\overline{\phantom{x}} : \F_{16} \to \F_{16} , x\mapsto \overline{x} = x^4 $$
be the nontrivial Galois automorphism of $\F_{16}$ with fixed field $\F_4$. 
Then the action of $h$ is given by
$$(x_1,y_1,x_2,y_2,\ldots , x_7,y_7) h = 
(\overline{y_1},\overline{x_1},\overline{y_2},\overline{x_2},\ldots , \overline{y_7},\overline{x_7})  .$$
Note that this action is only $\F_4$-linear.
In particular the fixed code of $\langle h \rangle$ is
$$C_1(h) = \{ (x_1,\overline{x_1},\ldots , x_7 , \overline{x_7}) \in C_1 \} $$
only an $\F_4$-linear code in $\F_{16}^{14}$.

\begin{cor}\label{defpi}
The code 
$X:=\pi (C_1(h)) := $
$$\{ (x_1,\ldots ,x_7) \mid (x_1,\overline{x_1},\ldots , x_7 , \overline{x_7}) \in C_1 \} \leq \F_{16}^7 $$
is an $\F_4$-linear trace-Hermitian self-dual code $X=X^{\perp}$ where 
$$X^{\perp }:= \{ v\in \F_{16}^7 \mid \sum _{i=1}^7 \trace_{\F_{16}/\F_4} x_i\overline{v_i} = 0 \mbox{ for all } x \in X \} $$
such that the minimal 5-weight of $X$ is at least 8.
Since $\dim _{\F_4}(X) = 7 = \dim _{\F_{16}} (C_1) $, 
the $\F_{16} $ linear code $C_1 \leq \F_{16}^{14}$ is obtained
from $X$ as 
$$C_1 = \Psi (X):= \langle (x_1,\overline{x_1},\ldots , x_7 , \overline{x_7}) \mid (x_1,\ldots , x_7) \in X \rangle _{\F_{16}} .$$
\end{cor}

\begin{rem}  {\rm 
If $\overline{\phantom{x}}:x\mapsto x^4$ denotes the Galois automorphism of $\F_{16}$ 
with fixed field $\F_4$, then $\wtfive(x) = \wtfive(\overline{x}) $ for all $x\in \F_{16}$.
Let $\xi $ denote a fixed element of order 5 in $\F_{16}^*$. 
Then $D_{10} = \langle \xi , \overline{\phantom{x}} \rangle $ acts $\F_4$-linearly 
on $\F_{16}$ and preserves the 5-weight and 
trace-Hermitian orthogonality.  }
\end{rem}

\begin{lemma}
An $\F_{4}$-linear code $X \leq \F_{16}^7$ with minimal $5$-weight at least $8$ is equivalent (under $D_{10} \wr S_7$) to a code
with generator matrix of the following type:
\begin{equation*}\label{eq:systematicForm}
\Gamma := \left(
\begin{array}{cccccc}
0   & 0   & 0   & 1 & \\
0   & 0   & 1   &   & \\
0   & 0   & \xi &   & \\
0   & 1   & 0   &   & \\
0   & \xi & 0   & a &  B \\
1   & 0   & 0   &   & \\
\xi & 0   & 0   &   & \\
\end{array} \right) , a \in (\F_4\cdot\xi)^6, B \in \F_{16}^{7 \times 3} 
\end{equation*}
We will call such a generator matrix \emph{systematic}.
\end{lemma}

\bew
The condition on the minimum 5-weight implies that there is at least one column with two $\F_4$-linearly independent entries.  Use the group action of $\GL_7(\F_4) \times \left( D_{10} \wr S_7 \right)$ to map this column to $(0, \ldots, 0, 1,\xi)^T$ and move the column to the front. Similar arguments can be applied to the derived code shortened at position $1$ and $\{1,2\}$, respectively.
\eb

\begin{theorem} \label{main3}
There is no extremal self-dual Type II code $C$ of length $72$ such that $\Aut(C)$ 
contains the dihedral group of order $10$.
\end{theorem}

\bew
Assume that there is such a code $C$ with $\Aut(C) \geq \langle g,h \rangle =G \cong D_{10}$.

Let $\Psi $ be the map from the $\F_4$-linear codes in $\F_{16}^7$ to 
the $\F_{16}$-linear codes in $\F_{16}^{14}$ 
from Corollary \ref{defpi}.
Let $\mathcal{C}_0$ be the 
list of 1920 codes of length 72 from Remark \ref{listeC0} and let
$\mathcal{X}$ denote a system of representatives of $D_{10}\wr S_7$
equivalence classes of  
 trace-Hermitian self-dual 
 codes $X \leq \F_{16}^7$ with minimal 5-weight at least 8. 
Then $$C\cong \varphi _0(C_0) \oplus \varphi _1 (\Psi(X)) \mbox{  
 for some } C_0 \in {\mathcal C}_0 \mbox{ and some } X \in \mathcal{X} .$$
For the proof of the theorem we summarize our construction method
 for all systematic generator matrices of $\F_{4}$-linear trace-Hermitian self-dual 
 codes $X \leq \F_{16}^7$ with minimal 5-weight at least 8 up to equivalence under 
$D_{10}\wr S_7$. 
 Furthermore, we restrict ourselves to these codes that might be extended by
 a binary code $C_0 \in \mathcal{C}_0$ such that $\varphi _0(C_0) \oplus \varphi _1(\Psi(X)) $
 has minimum distance $\ge 16$. 

The construction starts with the first row and iteratively adds a further row fulfilling the conditions
 on the systematic form. A backtracking approach is applied, whenever the condition 
on the 5-weight or self-orthogonality is violated, the code 
$\langle \Gamma_{1, \ast}, \ldots, \Gamma_{i-1, \ast} \rangle_{\F_4}$ is isomorphic to some
 other code already examined or there is no code $C_0$ such 
that $\varphi _0(C_0) \oplus \varphi _1(\Psi(\langle \Gamma_{1, \ast}, \ldots, \Gamma_{i-1, \ast} \rangle_{\F_4}))$ has minimum distance $\ge 16$.  

The following observations are used for speeding up the computations:
\begin{itemize}
 \item Each element in $\F_{16}^7$ is self-orthogonal under the
 trace-Hermitian inner product.
 \item We further know that each row $\Gamma_{i, \ast}$  must have minimum 5-weight at least 8. 
Since the 5-weight is not constant under scalar multiplication by elements $\mu \in \F_{4}^{\ast}$ we also have to test
$\wtfive(\mu \Gamma_{i, \ast} )$. This reduces the candidates for the first row to $3525$ vectors. There are $15705$ candidates for the other rows.
 \item The action of $\F_4^{\ast} \times \left( D_{10} \wr S_7 \right)$ partitions these $3525$ vectors into 6 orbits. It is sufficient to start with only one representative for each orbit.
 \item Similarly, for the $i$-th row it is sufficient to add only representatives under the action of the stabilizer of the code 
 $\langle \Gamma_{1, \ast}, \ldots, \Gamma_{i-1, \ast} \rangle_{\F_4}$.
 \item If some candidate $v$ for row $i$ is either not trace-Hermitian orthogonal to some preceding row $\Gamma_{j, \ast}, j \le i-2$
    or the minimum 5-weight of $\langle v, \Gamma_{1, \ast}, \ldots, \Gamma_{i-2, \ast} \rangle_{\F_4}$ is less than 8, we know that the corresponding permuted vector is not allowed to be a candidate for 
row $i+2$ and $i+4$, respectively.
 \item In the beginning there is a set ${\mathcal L}^{(0)} = {\mathcal C}_0$ of 1920 binary codes which may play the role of $C_0$. In each step $i$ of the iteration we may iteratively update this set by setting 
$$\begin{array}{l} 
{\mathcal L}^{(i)} := \{ C_0 \in {\mathcal L}^{(i-1)} : \\
d\left( \varphi_0(C_0) \oplus \varphi_1(\Psi(\langle \Gamma_{1, \ast}, \ldots, \Gamma_{i, \ast} \rangle_{\F_4})) \right) \ge 16\}.\end{array}$$ If ${\mathcal L}^{(i)}$ is empty we can skip this branch.
\end{itemize}

The test if there is another code already examined, which is isomorphic to the actual code 
is done by the calculation of unique orbit representatives by a modification of \cite{FeulnerCan}.
 This computation returns at the same time without any additional effort the stabilizer of
 $\langle \Gamma_{1, \ast}, \ldots, \Gamma_{i, \ast} \rangle_{\F_4}$ in $ D_{10} \wr S_7$. 
The computations have been performed in Magma \cite{MAGMA} and needed about 70 days CPU time.
 The number of non isomorphic candidates on level $i$ which appeared during our backtracking approach 
may be found in Table \ref{table:D10reps}. These numbers count $\F_{4}$-linear trace-Hermitian self-orthogonal codes which fulfill the condition on the given systematic form, the 5-weight and self-orthogonality. The test on the extendability by $C_0$ is executed after the isomorphism rejection. Hence, the numbers may vary for different backtracking approaches. For the remaining $4$ candidates at level $i=7$ the corresponding lists ${\mathcal L}^{(7)}$ of candidates for $ C_0$ are empty. 

In contrast to \cite{Feulner} applied in Section \ref{sec:Order_seven}, we preferred a row-wise generation of the generator matrix in this case, since this gives us
 the possibility to check the existence of a valid code $C_0 \in \mathcal{C}_0$.
\eb

\begin{table}[tb]
\begin{center}
\begin{tabular}{l|r}
$k$ & $ \begin{array}{l} \mbox{ Number of 
non isomorphic} \\ \mbox{ candidates for first $k$ rows} \end{array} $ \\ \hline
1 & 6\\
2 & 463 \\ 
3 & 4,885\\
4 & 856,804 \\ 
5 & 416,899 \\
6 & 306\\
7 & 4 \\\hline 
\end{tabular}
\caption{Computational Results for $D_{10}$}
\label{table:D10reps}
\end{center}
\end{table}


%

%




\end{document}